\magnification=\magstep1
\input amstex
\documentstyle{amsppt}
\pageheight{8.5truein}
\topmatter
\title {The maximal velocity of a photon}
\endtitle
\author   Avy Soffer \endauthor

\address Math Dept. Rutgers Univ. NJ, USA. \endaddress
\abstract

We estimate the probability of a photon to move faster than light,
under the Einstein dynamics which, unlike the wave equation or Maxwell wave dynamics,
has singular dispersion relation at zero momentum. We show that this
probability goes to zero with time, using propagation estimates
suitably multiscaled to control the contribution of low frequency
photons.
\endabstract
\endtopmatter
\document

 \head {Section 0. Introduction}
\endhead
\medskip

Let $F(\lambda > c)$ stand for a smoothed out characteristic
function of the interval $\lambda > c $. We say that a quantum
system, with a configuration coordinates $x$, obeys a maximal
velocity bound, if for its any evolution $\psi( x,t)$, localized in
a compact energy interval , there is a positive constant $c$ s.t.
$$
\| F \left( \frac{|x|}{t} >c \right) \psi( x,t) \| _{L^2} \to 0,
$$
as $|t| \to \infty.$ It is fairly easy to prove such bounds for
non-relativistic $N-$particle quantum systems (see \cite{SS, HSS}).
However, such bounds are still open for photons interacting with
such systems.
 The goal of the present paper is to prove such bounds for a model of a single photon with an effective interaction
 described by a potential $V(x)$. This model is given by
 the time-dependent Schr\"odinger equation

$$
i\frac {\partial \psi}{\partial t} = H\psi \quad \quad  \psi|_{t=0}
= \psi_0, \tag 0.1a
$$
with the  Schr\"odinger operator $H$ on $L^2 (\Bbb R^3)$,  given by
$$
H := (-\Delta)^{1/2} + V(x). \tag 0.1b
$$
This Hamiltonian is a simplified, scalar field version of the
restriction of the photon field in QED, to the one particle sector.
We think of the equation (0.1) as a model describing dynamics of a
single photon.
 Moreover, it is used as a laboratory for developing the methods to to study propagation with singular dispersion
 at zero energy, needed in tackling the full non-relativistic QED problem.
 Notice that the equation, without the interaction term, is
 relativistic invariant. Moreover, by multiplying the equation by
 $-i\partial_t -|p|$, we get the wave equation. However, since the
 initial data for the derived wave equation, is not localized:
 $$
 \dot \psi(0)=|p|\psi(0),
 $$
 we can not use the finite propagation property of the wave equation
 for this model.

 We make the following assumptions:

(i) $V(x) ,$  $  x\cdot \nabla V(x)$ are sufficiently regular and
 decay faster than $O(|x|^{-2})$ at infinity;

 (ii) $H$ has no zero energy resonances or zero energy
 eigenvalues.


The main result of this paper is the following theorem:

\proclaim{Theorem}
 Let $\psi(x,t)$ be the solution of the Schr\"odinger equation (0.1), and assume, furthermore,
 that $H$ satisfies the conditions (i) and (ii). If the initial data
 is localized in  energy in some compact interval, and is such that
 $ <x>^{1+\epsilon} \psi_0$ has a bounded $L^2$ norm, we have the
 following  maximal velocity bound:
 $$
\| F \left( \frac{|x|}{t} >R \right) \psi( x,t) \| _{L^2}= o(1) \tag
0.2
$$
for $ R \gg 1 $ \text{and} $|t| \to \infty.$
\endproclaim

 As was mentioned above, the proof of such a propagation estimates for general $N-$body Hamiltonians is quite
  elementary, see e.g. [Sig-Sof, HSS].
 See  also [BFS and cited references].
 However, the fact that
the one particle Hamiltonian of a massless photon is singular at zero
energy, prevents a direct extension of the general theory of
propagation estimates to the field, see however [BFSS, FGS].

The proof is based on the construction of propagation observables,
on each (dyadic) scale of the energy. Suppose that the length of the
momentum, $p=- i\nabla $, is localized in $ [2^{-n-1},2^{-n}],\quad
n\geq0$. Consider then the operator
$$
F(A/Rt2^{-n}>1)
$$
 defined through the spectral theorem.
$$A=-i/2(x\cdot \nabla +\nabla \cdot x), \tag 0.4 $$
 and $R>1$.

Then, we have the basic identity, derived from the Schr\"odinger
equation:
$$
\partial_{t} <\psi(t),F\psi(t)>= <\psi(t),\{i[H,F]+\partial_{t}
F\}\psi(t)>.\tag 0.5
$$
Here, $F$ stands for any self-adjoint operator, for which the
differentiation can be justified, for the chosen $\psi(t)$. To get a
useful propagation estimate, we use the idea of negative propagation
observables,[Sig-Sof]: Suppose $ F$ stands for a family of time
dependent operators, bounded from above (e.g., negative), and such
that its Heisenberg derivative, defined above, is positive, up to
integrable (in time) corrections. We will then obtain, upon
integration over time, the following propagation estimate:
$$
\int_{1}^{T} \|B\psi(t)\|^2 dt\leq C\|\psi(0)\|^2 .\tag 0.6
$$
Here, we use that the Heisenberg derivative, $D_{H}F$, is given by a
positive operator, plus integrable corrections:
$$
D_{H}F=i[H,F]+\partial_t F= B^*B + R(t).\tag 0.7
$$
In general, we need to use phase space operators which, while not
pseudo-differential, will have suitable phase space support. If we
choose for $F$, the function of $A$, defined above, as our
propagation observable, we need to estimate the commutator of $H$
with this $F$, suitably localized in some energy shell. One can then
use the following basic commutator expansion Lemma:
$$
\imath[H,F(A)]= F^{\prime}(A) \imath[H,A]+ R([[H,A],A]),\tag 0.8
$$
where the remainder is given by an explicit multiple integral,
involving the group generated by $A$, and, the double commutator
above, of $H$ with $A$. By the above expansion, we obtain, through
symmetrization, that the above propagation observable, has
Heisenberg derivative, that is smaller than:
$$
-\frac{1}{t}G_n^{2}(A/t)+R(t,n).\tag 0.9
$$
Here, $ G_n$ stands for a bump function of $\frac{A}{Rt2^{-n}}$
around $1$. We will get a useful estimate, if we can control the
remainder term, $R(t,n)$, by an integrable function of $t$, which is
also well behaved as $n$ tends to infinity. The leading term in the
expression (0.9), comes from the time derivative of $F$ term, of the
Heisenberg derivative. The first term in the Heisenberg derivative,
the commutator, has an expansion, beginning with:
$$
F^{\prime}(A/t)i[H,A]\frac{2^n}{Rt},\tag 0.10
$$
which can be shown to be much smaller than the leading term, by
using the energy localization around $2^{-n}$. To see that, we use
that the commutator $i[H,A]$ is bounded by $c|p|$, for some finite
constant $c$, by the use of the uncertainty principle and our decay
and regularity assumptions on $V$. Then, we need to show that energy
localization implies a similar momentum localization. This can not
be done using standard localization arguments, since the derivative
of the localization function grows like $2^n$! To this end, a
completely different argument is used: One proves that under generic
spectral assumptions on $H$, $HP_c(H)$, dominates a constant times
$P_c(H)|p|P_c(H)$. $P_c(H)$ stands for the spectral projection of
$H$ on its
 continuous spectral part.

 It then follows that:
$$
E_n(H)|p|E_n(H) \leq c2^{-n}E_n(H)^2.\tag 0.11
$$
Finally, we need to show that the remainder $ R(t,n)$, higher
commutator terms in the expansion of the Heisenberg derivative, are
integrable in $t$. This is the most involved step, without energy
localization the formal expression for the remainder is given by a
divergent integral. The way to estimate this last remainder term, is
to use the fact, that the group generated by dilations moves the
support of the energy localization functions. Consequently, the
integrals over the group actions is limited to finite domains. The
remainder term comes from the following expansion:
$$
\align E_n i [ |p|, \Phi_n ] E_n &= E_n i [ |p|^{1/2} , \Phi_n ]
|p|^{1/2} E_n
+ E_n |p|^{1/2} i [ |p|^{1/2} , \Phi_n ] E_n\\
&= E_n |p|^{1/2} F_n^{\prime}(A/t) |p|^{1/2} E_n \frac{1}{Rt} +2\Re
E_n |p|^{1/2} R_2 (A/t) E_n, \tag 0.12
\endalign
$$
with
 $$ \Phi_n \equiv E_{I_n}F(\frac{A}{tR2^{-n}}>1)E_{I_n}$$
 and with $R_2(A/t)$ given by
$$
R_2(A/t) = \frac{1}{2} \int d\lambda \hat F_n(\lambda) e^{i\lambda
A/Rt} \int_0^\lambda ds \int_0^s du \quad  e^{-iu A/Rt} \quad
\frac{1}{2} |p|^{1/2}\quad e^{iuA/Rt} (Rt)^{-2}$$$$ = \frac{1}{4}
\int d\lambda \hat F_n (\lambda) e^{i\lambda A/Rt} \int_0^\lambda ds
\int_0^s du \quad e^{-u /Rt} |p|^{1/2} \quad (Rt)^{-2}. \tag 0.13
$$
If we try to bound the expression for $R_2$, by taking the norm of
the first term on the rhs, as it was done in past works, we lose a
factor of $2^{-n/2},$ coming from localizing the $|p|^{1/2}$ factor.
On the other hand, we can not directly estimate the last expression
on the rhs of equation (0.13), since the integrand grows
exponentially, while $\hat F_n$ decays slower than exponential,
being the Fourier transform of a compactly supported function. To
this end, we use the fact that the dilation group, generated by $A$,
changes the support of functions of $|p|$, or $H$:
 $$ E_{I_n}(|p|)e^{i\lambda A}E_{I_n}(|p|)=0, \tag 0.14$$
 for $| \lambda |\ge \ln2.$
 Then, we use the mutual domination of $|p|,H$:
$$ P_c(H)H\le cP_c(H)|p|P_c(H)\le dP_c(H)H, \tag 0.15$$
for some positive constants $c,d$.

 Combining equations (0.14),
(0.15), we can then show that the integration on $\lambda$ is
limited to a compact domain, in equation (0.13).
 Collecting all of
the above, we get estimates of the form:
$$
E_n(H)\frac{1}{t}G_n^2(A/t)E_n(H) \in L^1(dt).\tag 0.16
$$
This estimate is then jacked up by the use of the propagation
observable $ \frac{A}{t}F_n(A/t),$ to obtain,
$$
E_n(H)F_n(A/t)E_n(H)\frac{1}{t} \in L^1(dt).\tag 0.17
$$
In the next step of the proof, we estimate, using the above, the
following operator:
$$
E_n(H) F(\frac{|x|}{t}>R)E_n(H).
$$
We write,
$F(\frac{|x|}{t}>R)=F(\frac{|x|}{t}>R)F_n(A/t)+F(\frac{|x|}{t}>R)\bar
F_n(A/t).$
$$
F_n+\bar F_n=1.
$$
The first term of the above decomposition, goes to zero, as time
goes to infinity, by the above propagation estimates, on $F_n(A/t).$

So, we need to show that the $\bar F_n$ term also goes to zero.

This is formally true , since it consists of a product of two
operators, with disjoint classical phase-space support, on the
energy shell $2^{-n}.$

 Again, the proof of this property
necessitates the use of new phase space localization arguments. In
the final step of the proof, we sum over all $n$, and in the process
we lose some powers of $2^{-n}$. These are compensated by requiring
the initial data to be localized in $x$, and by using that(up to 2)
negative powers of $|p|$, are bounded, up to a constant, by positive
powers of $|x|$.

\medskip

\head {\bf ACKNOWLEDGEMENTS} \endhead

I wish to thank I.M. Sigal for useful discussions. Part of this work
was done while the author visited the IHES, France. A. Soffer was
partially supported by NSF grant number DMS-0903651.

 \head {Section 1. Propagation Estimates}
\endhead
\medskip
Our goal in this section is to prove the following key propagation
estimate, as sketched in the introduction:
$$
\int_{1}^{T} \|F_n(A/t)E_n\psi(t)\|^2 \frac{dt}{t}\leq
C\|E_n\psi(0)\|^2 .\tag 1.1a
$$
and with

$$
F_n(A/t)\equiv F(\frac{A}{Rt2^{-n}}>1). \tag 1.1b
$$
$E_n$ stands for the operator $E_{I_n}(H)$.
 We use propagation
estimates, with the propagation observables
$$
    \Phi _n \equiv E_ {I_n} (H) F\left( \frac{A}{tR {2^{-n}} } > 1 \right) E_ {I_n} (H), \tag
    1.2a
$$
$$
\Phi _n \equiv E_ {I_n} (H) \frac{A}{t}F\left( \frac{A}{tR {2^{-n}}
}
> 1 \right) E_ {I_n} (H), \tag 1.2b
$$

 where, $    n = 0,1, \cdot \cdot
\cdot \infty , $ $I_n$ stands for the interval $[2^{-n-1} ,
2^{-n}]$, $E_ {I_n} $  is  the characteristic function of  $I_n $
and
 $A$ is the dilation generator,
$$
    A \equiv \frac{1}{2} (x \cdot p + p \cdot x). \tag 1.3
$$
Note that $ \frac{|x|}{t} \geq R $ on the support of $\Phi_n, $
 and $  R\leq \frac{|x|}{t} \leq 2R$ on the support of $E_nF^{\prime}E_n . $
Furthermore,
$$
    |\partial^j_\lambda F \left( \frac{\lambda}{Rt2^{-n}} > 1 \right)| \sim \left( \frac{2^n}{Rt} \right) ^j . \tag 1.4
$$

The main propagation estimate is based on showing that:
$$
\align
& E_{I_n} i \left[ |p| + V, F \left( \frac {A}{Rt2^{-n}}
> 1 \right) \right] E_{I_n}\\
&=E_n \left \{ i |p|^{1/2} \left[ |p|^{1/2} , F \right] + i\left[
|p|^{1/2}, F \right] |p|^{1/2} + i[V,F] \right \}E_n \\ &= E_n \left
\{ |p|^{1/2} \tilde{F}^2 |p|^{1/2} + |p|^{1/2} \frac
{2^{2n}}{R^2t^2} Q_2(A/t) |p|^{1/2} \right \}E_n+O(t^{-2})  ,  \tag
1.5a
\endalign
$$
where $\tilde{F}^2 \sim \frac{2^n}{Rt} F^{\prime}$ \quad  and,
$E_nQ_2(A/t)E_n$ is of order 1 for $t \ge 2^n/R,$ and of order
$2^{-n},$ for $t \le 2^n/R.$
 Then, we show that
$$    E_n |p|^{1/2} \tilde{F}^2 | p|^{1/2} E_n \lesssim (c/Rt) E_n
G_n^2 E_n,\tag 1.5b $$ with
$$G_n^2=F(\frac{A}{Rt2^{-n}}\sim 1). $$

Letting $ F_n(\lambda)=F(2^{n}\lambda \ge 1)$ and using that
$-\frac{A}{t}F_n^{\prime} \sim -R2^{-n}F_n^{\prime}, R\gg1 $, we
obtain
$$
\align
\int _1^T\partial_t < \psi_t, E_{I_n}(H)F_n(\frac{A}{Rt}) &E_{I_n}(H)\psi_t > dt = \int_1^T  <\psi_ti[H,\Phi_n]\psi_t >dt
 \\
&+
2^{n}R^{-1}\int_1^T<\psi_t,E_{I_n}(H)(\frac{-A}{t^2})F_n^{\prime}E_{I_n}(H)\psi_t>dt.\tag
1.6
\endalign
$$
Assuming that (1.5) holds, using (1.6) we can then prove:

\proclaim{Theorem 1.1}
 Under the previous assumptions on $H$, and
assume that (1.5) holds, we have the following propagation
estimates:
$$
\int_1^T \|F_n^{\prime}(A/t)E_n\psi(t)\|^2 \frac{dt}{t} \le
c(R)\|E_n\psi(0)\|^2\tag 1.7a
$$

$$
\align
 \int_1^T \|F_n(A/t)E_n\psi(t)\|^2& \frac{dt}{t} \\+<\psi(T),E_n
\frac{A}{T}F_n(A/T)E_n\psi(T)> &\le c(R)
 \|
<A>^{1/2}E_n \psi(0)\|^2 \tag1.7b
\endalign
$$

\endproclaim
\demo{proof}

 To prove (1.7a), we observe, that by the estimate
(1.5), combined with Proposition(2.4), each power of $|p|$
contributes a factor of $2^{-n}$, the $\tilde F$ term is bounded by

 $(c/Rt)F^{\prime}, $ which is dominated by the integrand of the
 second term in equation (1.6), for $R$ large enough.
 The $Q_2$ term contribution to the commutator in equation (1.6), is
 bounded by our assumptions , following estimate (1.5):
 $$
 \|E_n|p|^{1/2}\| \le c2^{-n/2},
 $$
 by Proposition (2.4),which eliminates one power of $2^n$; then,
 integrating $1/t^2$ over the region of $t\ge 2^n/R$, eliminates a
 power of $2^n$. The integral over $t \le 2^n/R$, is also bounded,
 since $Q_2$ is assumed to be smaller than $c2^{-n}.$
 To prove (1.7b), we use the propagation observable (1.3b).
The proof is similar to the first case: The time derivative term,
which contributes the second term on the rhs of (1.6), is , as
before the dominant term. The $Q_2$ term is controlled as before
(the only difference is that now $\hat F$ is replaced by
$\partial_{\lambda} \hat F(\lambda)$, in the estimates on $Q_2$. As
for the commutator term, we have now
$$
i[H,\Phi_n]=E_n\frac{1}{t}i[H,A]F_n(A/t)E_n+
E_n\frac{A}{t}F^{\prime}_n(A/t)i[H,A]\frac{2^n}{Rt}E_n+
E_nQ_2E_n.\tag 1.8
$$
The $Q_2$ term is estimated as before. The second term on the rhs of
91.8) is also estimated as before, since the factor $A/t \sim
2^{-n}/R,$ on support $F^{\prime}.$ However, the first term is
unbounded, so the bound for $t\le 2^n/R,$ Proposition (C), does not
apply. Instead, we break the $F$ to two parts, one decaying at
infinity,and sharply localized in Fourier space, and one constant at
infinity. The first part is estimated as before, and the second part
is directly controlled by elementary size estimates, see Proposition
(D).$\blacksquare$
\enddemo

 There are new difficulties in
completing this argument, compared with the usual case, without
dyadic energy localization.

 First, we
need to minimize the number of powers of $2^n$, coming from
expanding the function $F_n$.
 Then, we need to trade positive
powers of the momentum( derivative operator) $p =-i \nabla$,  for powers of
$2^{-n}$.

Finally, to control the remainder term in the Commutator Expansion
Lemma, the $Q_2$ term, (or $R_2$ term), we need to commute the
derivative through the dilation group, which produces exponentially
large factors.

 The way out of these problems involves the following
arguments.
To limit the integrations in the remainder term $R_2$, we notice that,
the dilation group moves the dyadic energy interval, away from its
original support.
 Hence, for large enough value of the group parameter, $\lambda$,
the fact that our propagation observable is localized on the dyadic
interval, from both sides, gives an extra decay, that cancels the
exponential growth factor. This is shown in detail in the subsection
"The term $R_2$".

 To get the $2^{-n}$ factor from the momentum $p$, we prove some
propositions about the properties of the operator $H$, which might
be of independent interest.(see Proposition (2.4))
 Specifically, we show, that in three (and higher)
dimensions, if there are no zero energy resonances and eigenvalues,
then $H$ and $|p|$ dominate each other, up to a multiplicative
constant, on the continuous spectral subspace of $H$.

 These estimates are the key to getting the
right minimal powers of $2^n$, from the various propagation estimates
and phase space localizations.
\smallskip
To prove the necessary estimates, to fill in the details of the
above theorem, and the assumptions on $Q_2$, we estimate the
Heisenberg derivative of the observables defined in (1.3), on the
time intervals $t\ge 2^n/R,$ and $t \le 2^n/R.$ This is the content
of the following four Propositions A-D.

 We begin by deriving and
estimating the $R_2$ term:
\medskip

\head {\bf The $R_2(A/t)$ term.}\endhead
\medskip
\proclaim{Proposition A}

For $\Phi_n$ as in (1.3a), we have:
$$
\int_{2^n/R}^T <E_n|p|^{1/2}Q_2|p|^{1/2}E_n> dt \le
 \frac{c}{R}\|E_n\psi(0)\|^2. \tag 1.9
$$

\endproclaim
\demo{Proof}

Direct application of the commutator expansion lemma gives:

$$
 i \left[ |p|^{1/2}, F_n(A/Rt) \right]
 = \int d\lambda \hat F_n (\lambda) e^{i\lambda A/Rt} \frac{1}{Rt} \int_0^\lambda e^{-isA/Rt}
  i\left[ A, |p|^{1/2} \right] e^{isA/Rt}ds $$$$
= \frac{1}{2} F_n^{'}(A/t) \frac{1}{Rt} |p|^{1/2} + \tilde R_2 (A/t)
|p|^{1/2} = \frac{1}{2} F^{\prime}_n(A/t) \frac{1}{Rt} |p|^{1/2} +
R_2(A/t), \tag 1.10a
$$
where we used that
$$
    i \left[ A, |p|^{1/2} \right] = p \cdot \nabla_p |p|^{1/2} = \frac{1}{2} |p|^{1/2}. \tag
    1.10b
$$

$$
R_2(A/t) = -\frac{1}{2} \int d\lambda \hat F_n(\lambda) e^{i\lambda
A/Rt} \int_0^\lambda ds \int_0^s du \quad  e^{-iu A/Rt} \frac{1}{2}
|p|^{1/2}\quad e^{iuA/Rt} (Rt)^{-2}$$$$ = \frac{1}{4} \int d\lambda
\hat F_n (\lambda) e^{i\lambda A/Rt} \int_0^\lambda ds \int_0^s du
\quad e^{-u /Rt} |p|^{1/2}(Rt)^{-2}. \tag 1.11
$$
In general, this integral blows up at infinity, due to the fact that $e^{u/Rt}$ grows exponentially fast, while
  $\hat{F}_n(\lambda)$  decays faster than any polynomial, but not exponentially,
  since $\hat {F}_n(\lambda)$ is the Fourier transform of a compactly supported function.
$$
\align
E_n i [ |p|, \Phi_n ] E_n &= E_n i [ |p|^{1/2} , \Phi_n ] |p|^{1/2} E_n
+ E_n |p|^{1/2} i [ |p|^{1/2} , \Phi_n ] E_n\\
&= E_n |p|^{1/2} F_n^{\prime}(A/t) |p|^{1/2} E_n \frac{1}{Rt} + E_n
|p|^{1/2} R_2 (A/t) E_n, \tag 1.12
\endalign
$$
with $R_2(A/t)$ given by (1.11).
$$
\align
E_n |p|^{1/2} R_2 (A/t) E_n &= E_n |p|^{1/2} \tilde{E}_n(|p|) R_2 (A/t) \tilde{E}_n(|p|) E_n\\
& \quad + E_n O(2^{-n}n) \bar {E}_n (|p|) |p|^{1/2} R_2 (A/t) \tilde{E}_n (|p|) E_n\\
& \quad + E_n |p|^{1/2} \tilde {E}_n (|p|) R_2 (A/t) \bar {E}_n (|p|) O (2^{-n}n) E_n \\
&\equiv J1 + J2 + J3. \tag 1.13
\endalign
$$
$$
\bar {E}_n \equiv 1 - \tilde {E}_n .
$$
In our case,  $ 1 - \tilde {E}_n = \tilde {E} (|p| \leq 1 ) - \tilde {E}_n$.
$\tilde {E}_n$  stands for smoothed $E_n$ function, and where we used proposition 2.2(d).

$$
\align \int_1^T J_1 dt&=\int_1^T E_n\|p|^{1/2}\tilde {E}_n (|p|) R_2(A/t)\tilde {E}_n (|p|) E_n dt \\
&=\int_1^T \frac{dt}{R^2t^2} E_n |p|^{1/2}\tilde {E}_n (|p|) Q_2
(A/t)\tilde {E}_n (|p|)|p|^{1/2} E_n\\
&= \int_1^{c2^n} \frac{dt}{R^2t^2} E_n |p|^{1/2}\tilde {E}_n (|p|) Q_2 (A/t)\tilde {E}_n (|p|)|p|^{1/2} E_n \\
&+ \int _{c2^n}^T \frac {dt}{R^2t^2} E_n |p|^{1/2} \tilde {E}_n
(|p|)Q_2(A/t)\tilde {E}_n (|p|)|p|^{1/2}E_n .\tag 1.14
\endalign
$$
If $T \leq c2^n$ , then the second term on the r.h.s of (1.14) is
zero.
\smallskip
{\bf \quad \quad \quad \quad Estimating J1}
\medskip

 Using
proposition (2.2c), it follows that the $\lambda$ integration (and
therefore the other integrations)is limited to
 $$
    |\lambda| \leq Rt \ln2. \tag 1.15
$$
Hence,
$$
J1 =\frac{c}{(Rt)^2}
E_n (H) |p|^{1/2} \tilde {E}_n(|p|)
\int_{|\lambda| \leq Rt \ln2} d\lambda\hat{F}_n (\lambda)
e^{i\lambda A/Rt}
 $$
$$
\times\int_0^\lambda ds \int_0^s du e^{iu A/Rt} |p|^{1/2} e^{-iu
A/Rt}   \tilde {E}_n(|p|) E_n(H). \tag 1.16
$$

$$
\align
J1= \frac{c}{(Rt)^2}
&E_n |p|^{1/2} \tilde {E}_n (|p|) \int_{|\lambda| \leq Rt \ln2} \hat {F}_n (\lambda) e^{i\lambda A/Rt}
 \int_0^\lambda ds \int_0^s e^{-u/Rt}du
 |p|^{1/2} \tilde{E}_n (|p|) E_n(H) \\
&=\frac{1}{(Rt)^2}O( E_n 2^{-n/2} \int_{|\lambda| \leq Rt \ln2} |\lambda^2 \hat {F}_n (\lambda)|
\quad d\lambda \quad 2^{-n/2} E_n )\\
&= \frac{1}{(Rt)^2}O( E_n 2^{-n} (Rt)^2 E_n ).
\endalign
$$
Hence,
$$
    \int_1^{2^n/R}  J1 \quad dt \leq O(\frac{1}{R}).
$$
If  \quad $Rt > 2^n$, we use instead, that
$$
    \int |\lambda^2 \hat {F}_n (\lambda)| \quad d\lambda \leq 2^{2n},
$$
so that,
$$
    J1 = \frac{1}{(Rt)^2}O ( E_n 2^n E_n ), \quad \text {and then},
$$
$$
    \int_{2^n /R} ^T  J1 dt = O(\frac {1}{R}).\tag 1.17
$$

\medskip
 {\bf \quad \quad \quad \quad The estimates of $J2$ and $J3$.}
\medskip
Consider the region \quad $Rt > 2^n $.
 The integrand to estimate, which  is

$\left ( |p|_s \equiv e^{-iAs/Rt}|p| e^{iAs/Rt} \right),$ can be
written as,
$$
(Rt)^{-2 } \hat {F}_n(\lambda) \tilde{J}(\lambda,s,u),
$$

$$
    \tilde {J}(\lambda,s,u) \equiv E_n(H) |p|^{1/2} \tilde{E}_n|(p|) e^{i\lambda A/Rt} |p|_u ^{1/2}
     \bar {E}_n (|p|) E_n(H),
$$
and adjoint of such term.

First, we decompose the region of integration $\lambda$ to:
$$
    \frac {|\lambda|}{Rt} > m\
    \text {and }  \frac {|\lambda|}{Rt} \leq m,\
    m > \ln2.
$$
For $\frac {|\lambda|}{Rt} > m$ we commute $\tilde {E}_n (|p|)$ through $e^{i\lambda A/Rt}$, to get,
$$
\tilde{J}(\lambda, s, u) = E_n(H) |p|^{1/2} \tilde {E}_n (|p|) e^{i
\lambda A/Rt} |p|_{u} ^{1/2} \bar{E}_{\bar{n}(\lambda)} (|p|)E_n(H),
$$
for some $\bar{n}(\lambda) \neq n$ , (since $\frac{|\lambda|}{Rt} >
\ln2$, by assumption), and so $\bar {E}_n E_{\bar{n}(\lambda)} =
E_{\bar{n}(\lambda)}.$
Then,
$$
 \bigg| \int_{2^n/R} ^T \frac{dt}{R^2t^2}
\int_{-\infty}^{\infty} \hat{F}_n (\lambda)\int_{s=0}^{\lambda}
\int_{u=0}^{s} \tilde{J}(\lambda, s,u) d\lambda ds du \bigg|
$$

$$ \leq \frac{1}{R^2} 2^{-n} R
\sup_t  \int_{\frac{|\lambda|}{t} > m}|\hat{F}_n(\lambda) \lambda ^2 | d\lambda \cdot
2^{-n/2} \cdot 2^{-n} $$$$ \leq \frac{1}{R} 2^{-n/2} 2^{-2n} O(|m2^n|^{-k}).
$$
In the last estimate, the factor $2^{-n} R$ comes from the $t-$ integration.
The factor $2^{-n/2}$ comes from $|p|^{1/2} \tilde {E}_n (|p|)$.
 This factor is missing in the adjoint terms.
  Another factor, $2^{-n}$ , comes from applying proposition 2.2 to the product
   $E_{\bar{n}(\lambda)} (|p|) E_n (H)$.

   Finally, we note that, by construction, $\hat {F}_n (\lambda)$ is vanishing faster than any polynomial at infinity,
    and is approximately a constant, of order $R2^{-n}$, on an interval of size $2^n/R \quad$ around the origin.
     Since $|\lambda| > mRt > m2^n$ , the $O(|mt|^{-k})$ bound, $\forall k,$ follows.

      Next, we consider the region $\frac {|\lambda|}{t} \leq m$.
Now, $\frac {|\lambda|}{t} \leq m \quad \Rightarrow \quad \frac {|s|}{t}, \frac{|u|}{t} \leq m$.
Therefore $\quad e^{|s|/Rt}, e^{|u|/Rt} \leq e^m \quad$,
 so, $\quad |p|_s \leq e^m |p| \quad $.
So, we pick up a factor of $2^{-n}$ from $|p|^{1/2}$ factors, and the integration of $u, s, t $ gives:
$$
    \int_{\frac {|\lambda|}{Rt} \leq m} |\lambda^2 \hat{F}_n (\lambda) | d\lambda \leq c(2^{2n}/R^2) e^m.
$$
Hence,
$$
    \bigg| \int_{2^n/R} ^T \frac {dt}{R^2t^2}\int \int \tilde{J}(\lambda, s, u) \hat{F}_n(\lambda) d\lambda ds du
    \bigg|
    \leq c \frac {e^m}{R^2} 2^{-n} R 2 ^{-n} (2^{2n} / R^2) 2^{-n},
$$
where we gain an extra $2^{-n}$ from $E_{\bar n} E_n$ and $|p|$
each.

\smallskip

 {\bf Large $|p|$ or $H$.}

\medskip

  Since the momentum operator is
unbounded, we need to bound it, under the integrals, in the
definition of $R_2$, for large values of $|p|$, or $H$.
 To this end,
we replace $|p| \longrightarrow Hg(H \leq 10)$ ,and then, we need to
work with:
$$
-|p| \longrightarrow i[A,Hg] \quad \text{and} \quad i[A, [A,Hg]].
$$
$$
\align
-i[A,Hg] &= (|p| + x \cdot \nabla V) g - Hi[A,g(H)] \\
&= g(|p| + x \cdot \nabla V) - i[A,g(H)] H.
\endalign
$$
This change is not difficult to handle as before.

In cases we only used the boundedness of these commutators, it is
straight-forward, since both $g(H)|p|$ and $i[A,g(H)]H $ are
bounded.

In case we need to use a factor of $|p|$ (or $|p|^{1/2}$) to get
extra smallness factor $2^{-n}$, it is done as before, since the
$g(H)$ factor does not change $E_n(H)$. The $x\cdot \nabla V$ term
is better, since it is bounded by $c|p|^2.$ $\blacksquare$

\enddemo
\smallskip
\head {\bf The region $0 \leq t \leq 2^n / R.$} \endhead

\medskip

\proclaim{Proposition B}

 For $\Phi_n$ as in (1.3a), we have:
$$
\int_1^{2^n/R} \|G_n(A/t)E_n\psi(t)\|^2\frac{ dt }{t}\le
 c\|E_n\psi(0)\|^2. \tag 1.18
$$
Here, $G_n$, is a bump function of $A/t$ around $2^{-n}/R.$
\endproclaim
\demo{Proof}

Now we have,
$$
\frac{d}{dt}\left( \psi(t), \Phi_n \psi(t)\right) = \left( \psi(t),
i[H,\Phi_n] \psi(t)\right) + \left(\psi(t), \frac{d\Phi_n}{dt}
\psi(t)\right),
$$
$$
\left(\psi(t),\frac{d\Phi_n(t)}{dt} \psi(t)\right) =
\left(E_n\psi(t), t^{-1}
F_n^{'}(A/t)\left(-\frac{A}{t}\right)E_n\psi(t)\right)
$$
$$
\leq -\frac{1}{t}\left(E_n\psi, 2^{-n}RF_n^{'}(A/t)E_n\psi\right)
\equiv -\frac{1}{t}\left(E_n\psi,\tilde{F}_n^2(A/t)E_n\psi\right),
$$
where,
$$
\align
&|\tilde{F}_n^2(A/t)| \lesssim 1 \quad \text{and} \quad \tilde{F}_n^2 \geq 0 \\
&\tilde{F}_n^2(A/t) \simeq 1 \quad \text{for} \quad \frac{A}{t} \sim R2^{-n} \\
&\tilde{F}_n^2(A/t) = 0 \quad \text{for} \quad \frac{A}{t} \nsim
R2^{-n}.
\endalign
$$
$$
\align
\left(\psi(t),i[H,\Phi_n]\psi(t)\right) &= \left(\psi(t),i(HE_nF_nE_n - E_nF_nE_nH)\psi(t)\right) \\
&= i\left(E_n\psi(t),(HE_nF_nE_n - E_nF_nE_nH)E_n\psi(t)\right) \\
&\leq 2\cdot 2^{-n} ||E_n\psi(t)||^2.
\endalign
$$
Hence,
$$
\align
&\int_0^{2^n/R}\biggl\{\Bigl(\psi(t),i[H,\Phi_n]\psi(t)\Bigr) + \Bigl(\psi(t),\frac{d\Phi_n}{dt}\psi(t)\Bigr)\biggr\}dt
 \\
&\leq -\int_0^{2^n/R}\frac{dt}{t}\Bigl\| \tilde{F}_n(A/t)E_n\psi(t)\Bigr\|^2 +
 2\Bigl\|E_n\psi(t)\Bigr\|^2 \int_0^{2^n/R}2^{-n}dt \\
&= - \int_0^{2^n/R} \frac{dt}{t} \Bigl\| \tilde{F}_n (A/t) E_n \psi(t) \Bigr\|^2 +
 \frac{2}{R} \Bigl\|E_n\psi(t)\Bigr\|^2.
\endalign
$$

Therefore,
$$
\align
\int _0 ^{2^n/R} \frac{dt}{t} &\Bigl\| \tilde{F}_n (A/t) E_n \psi (t)\Bigr\|^2 + \Bigl( \psi (2^n/R), F_n(RA/2^n) \psi(2^n/R) \Bigr)\\
&\leq \Bigl( E_n \psi(0), F_n(A,t=0) E_n \psi(0) \Bigr) +
\frac{2}{R} \Bigl\| E_n \psi(t) \Bigr\|^2.
\endalign
$$

$\blacksquare$

\enddemo
\smallskip
 \centerline {\bf Improved Decay}
\medskip
\proclaim {Proposition C}

Under the assumptions as the above Propositions (A,B), we have the
following propagation estimate,
$$
\align
 \int_{2^n/R}^T \|F_n\left(\frac{A}{Rt2^{-n}}>1\right
)E_n\psi(t)\|^2& \frac{dt}{t} \\ +
<\psi(T),E_n\frac{A}{T}F_n\left(\frac{A}{RT2^{-n}}>1\right
)E_n\psi(T)> & \le c(R)\|E_n <A>^{1/2}E_n\psi(0)\|^2,\tag 1.19
\endalign
$$
for all $R$ sufficiently large.
\endproclaim

\demo{Proof}
 We can use $E_n \frac{A}{t} F_n (A/t) E_n = \Phi_n $ .
The estimate of $R_2(A/t)$ is the same as before, so for $t\geq 2^n$
, it is done as before.The main change is that now $\hat F(\lambda)$
is replaced by $\partial_{\lambda} \hat F(\lambda),$, and a similar
proof applies. The first term in the commutator expansion of
$i[H,\Phi_n],$ has an extra, positive term, which is however
integrable over time by the Previous propositions (A,B), since it is
supported (in phase space) on the support of $F_n^{\prime}.$
$\blacksquare$
\enddemo

For $0\leq t\leq 2^n$, since $\frac{A}{t}F_n$ is not bounded, the
proof is different.

\proclaim{Proposition D}
$$
\align  \int_1^{2^n/R} \|F_n\left(\frac{A}{Rt2^{-n}}>1\right
)E_n\psi(t)\|^2& \frac{dt}{t} \\+
<\psi(2^n/R),E_n\frac{A}{2^n/R}F_n\left(A>1\right )E_n\psi(t)> & \le
c(R)\|E_n <A>^{1/2}E_n\psi(0)\|^2,\tag 1.19
\endalign
$$
for all $R$ sufficiently large.
\endproclaim
\demo{Proof}

 We write,
$$\frac{A}{t}F_n = F_{n,1} + F_{n,2},$$ where,
$$F_{n,1} =i \int e^{-\lambda ^2} \partial_\lambda \hat{F}_n(\lambda) e^{-i(A/t)\lambda} d\lambda ,$$
and
$$  \quad F_{n,2} \equiv (A/t)F_n - F_{n,1}. $$
Then, $F_{n,2}$ is bounded, and the previous proof applies, while
$F_{n,1} \equiv \frac{A}{t}G_n$, with $ G_n$ smooth, approaching a
constant at infinity. Then,
$$
\align
i[H,F_{n,1}] &= \frac{1}{t}i[H,A]G_n + \frac{A}{t}i[H,G_n] \\
&= \frac{1}{t}\bigl[|p|+\tilde V(x)\bigr]G_n +\frac{A}{t}\int \hat{G}_n(\lambda) e^{-i\lambda A/t}
\int_0^\lambda e^{isA/t} [H,A/t] e^{-isA/t} ds d\lambda .
\endalign
$$
The first term is bounded by \quad $2^{-n}\|G_n\|$ \quad on support
of $E_n$, and therefore the integral over \quad $1 \leq t \leq2^n$
 is bounded by $O(1)$.
 The second term is,
$$
\align
\frac{c}{t}\int\partial_\lambda \Bigl(\lambda \hat{G}_n(\lambda)\Bigr) e^{-i\lambda A/t} &\frac{1}{\lambda}
 \int_0^\lambda e^{-s/t} \Bigl[ |p|+\tilde {V}(x,s)\Bigr]ds \\
&+\frac{c}{t}\int \hat{G}_n(\lambda) e^{-i\lambda A/t} \frac{1}{\lambda} \int_0^\lambda e^{-s/t}
 \Bigl[|p|+\tilde{V}(x,s)\Bigr]ds.
\endalign
$$
$$ \frac{1}{\lambda}\int_0^\lambda e^{-s/t} ds = -\frac{t}{\lambda}
e^{-s/t} |_0^\lambda = \frac{t}{\lambda} \bigl( e^{-\lambda /t} -1
\bigr) $$
  and, for  $\frac{\lambda}{t} \ll 1, \quad
e^{-\lambda/t} -1 \sim -\frac{\lambda}{t} + \frac{1}{2}
\frac{\lambda ^2}{t^2}$,

 so the first term contributes $
\int \hat{G}_n (\lambda) e^{i\lambda A/t} d\lambda = O(1),  $
and the second term is bounded by \quad $c\int|\lambda
\hat{G}_n(\lambda)| d\lambda /t$.

Furthermore, there is a factor of $2^{-n}$, coming from
 $|p| + \tilde{V}(x,s) \leq |p| + c|p|^2$.

Here, $ \tilde{V}(x,s)=e^{isA/t}\tilde{V}(x)e^{-isA/t}.$
$\blacksquare$
\enddemo

 \head {Section 2. Auxiliary Identities and Inequalities}
\endhead
\medskip
\proclaim {Lemma 2.1}
Assume $H=|p|+V$ and $|V| <
\frac{1}{2r}$. Then,
$$
\Bigl\||p|E_{I_n}(H)f\Bigr\|+\Bigl\|V E_{I_n}(H)f\Bigr\| \leq c2^{-n} \Bigl\|E_{I_n} f\Bigr\| .\tag 2.1
$$
\endproclaim
\demo{Proof}
For $\|f\|=1$;
$$
\align
\Bigl(f,E_{I_n}\bigl(|p|^2+V^2\bigr)E_{I_n}f\Bigr) &= \Bigl(f,E_{I_n}(H)\{H^2-V|p|-|p|V\}E_{I_n}(H)f\Bigr) \\
& \leq
2^{-2n}\Bigl\|E_{I_n}f\Bigr\|^2+2\Bigl\|VE_{I_n}(H)f\Bigr\|\Bigl\||p|E_{I_n}(H)f\Bigr\|.
\endalign
$$
Let $a=\Bigl\||p|E_{I_n}f\Bigr\| \quad b=\Bigl\|VE_{I_n}f\Bigr\|$.
 Then, $\quad b<(1-\delta)\Bigl\||p|E_{I_n}f\Bigr\|$ by
the uncertainty inequality(in 3-dimensions or higher), and so, \quad
$$
\align
\Bigl(f,E_{I_n}\bigl(|p|^2+V^2\bigr)E_{I_n}f\Bigr) &= (a^2+b^2-2ab) + 2ab\\
&\geq\delta ^2\Bigl\||p|E_{I_n}f\Bigr\|^2 + 2ab.
\endalign
$$
It follows  that, $\quad \delta ^2 \Bigl\||p|E_{I_n}f\Bigr\|^2 \leq
2^{-2n} \Bigl\|E_{I_n}f\Bigr\|^2$ .$\blacksquare$
\newline
\enddemo
\proclaim{Proposition 2.1} If $|p| \lesssim H$,  then (2.1)
holds.
\endproclaim
\demo{Proof}
 When $|p| \leq mH$, we have that, by
the spectral theorem,
$$
\frac{1}{H}\leq\frac{m}{|p|} \Longrightarrow \Bigl\|\frac{1}{H^{1/2}}f\Bigr\|_{L^2}
\leq m^{1/2}\Bigl\|\frac{1}{|p|^{1/2}}f\Bigr\| .\tag 2.2
$$
Hence,
$$
V(x)\frac{1}{H} = V(x)\left(\frac{1}{|H|} - \frac{1}{|p|}\right) + V(x)\frac{1}{|p|}.
$$
$$
\Bigl\|V(x)\frac{1}{|p|}\Bigr\| = \Bigl\|V(x)r\frac{1}{r}\frac{1}{|p|}\Bigr\|
\leq \Bigl\|V(x)\frac{1}{r}\Bigr\|_\infty 2\Bigl\||p|\frac{1}{|p|}\Bigr\| <\infty,
$$
$$
\align
V(x)\left(\frac{1}{H}-\frac{1}{|p|}\right) &= -V(x)\frac{1}{H}V(x)\frac{1}{|p|} \\
&= -V(x)\frac{1}{H}\frac{1}{r}rV(x)r\frac{1}{r}\frac{1}{|p|} \\
&= -V(x)r\frac{1}{r}\frac{1}{H}\frac{1}{r}r^2V(x)\frac{1}{r}\frac{1}{|p|}.
\endalign
$$
Since,
$$
\frac{1}{r}\frac{1}{H}\frac{1}{r}=\frac{1}{r}\frac{1}{H^{1/2}}\frac{1}{H^{1/2}}\frac{1}{r}
=\left(\frac{1}{H^{1/2}}\frac{1}{r}\right)^*\left(\frac{1}{H^{1/2}}\frac{1}{r}\right),
$$
we get that,
$$
\align
\Bigl\|V(x)\left(\frac{1}{H}-\frac{1}{|p|}\right)\Bigr\|
&\leq \Bigl\|\frac{1}{H^{1/2}}\frac{1}{r}rV(x)\Bigr\|\Bigl\|\frac{1}{H^{1/2}}
\frac{1}{r}r^2V(x)\frac{1}{r}\frac{1}{|p|}\Bigr\| \\
&\leq m\Bigl\|\frac{1}{|p|^{1/2}}\frac{1}{r}\bigl(rV(x)\bigr)\Bigr\|\Bigl\|\frac{1}{|p|^{1/2}}\frac{1}{r}(r^2V)
\Bigr\|\Bigl\|\frac{1}{r}\frac{1}{|p|}\Bigr\| \\
&< mC_V,
\endalign
$$
where $C_V$ is a constant, depending on the $L^\infty$ norm of $V,rV,r^2V$.\newline
We therefore conclude that,
$$
|p|\frac{1}{H} = \bigl(|p|+V-V\bigr)\frac{1}{H} = H\frac{1}{H}-V\frac{1}{H}=1-V\frac{1}{H}
$$
is also bounded.
Finally, we have that,
$$
|p|E_{I_n}(H)=|p|\frac{1}{H}HE_{I_n}(H)=O(1)2^{-n}E_{I_n}(H).\quad
\quad \blacksquare
$$
\enddemo
Now we prove some useful identities.
 \proclaim {Lemma 2.2 }
\medskip
\noindent (i)\quad $r^2(-\Delta)=A^2+L^2-iA-3/4$,  \quad
(dimension$=3$).
\newline
(ii) \quad $e^{i\lambda A}|p|^\alpha e^{-i\lambda A}=e^{-\alpha
\lambda}|p|^\alpha$ ; \quad $A\equiv
\frac{1}{i}(r\partial_r+\frac{3}{2})$. \newline (iii)  \quad
$e^{i\lambda A} r^\alpha e^{-i\lambda A}=e^{\alpha \lambda} r^\alpha
. $ \newline (iv)    \quad
$\frac{1}{|p|}A\frac{1}{r}=\frac{1}{|p|}\Bigl(i|p|\frac{\partial}{\partial
|p|}-\frac{3i}{2}\Bigr)\frac{1}{r}=\frac{1}{|p|}\Bigl(ip\cdot
\frac{\partial}{\partial p} -\frac{3i}{2}\Big)\frac{1}{r}=O(1).$
\newline (v) \quad $\Bigl\|\frac{1}{|p|}\frac{1}{r}\psi\Bigr\| \leq
2\|\psi \|. $\endproclaim

  \demo{Proof}
  \newline (i) \quad
$r^2(-\Delta)=r^2\Bigl(-\partial_r^2-\frac{2}{r}\partial_r
+\frac{L^2}{r^2}\Bigr)=-r^2\partial_r^2-2r\partial_r+L^2$
\newline
$=-2r\partial_r+2r\partial_r-r\partial_r^2r+L^2=-r\partial_r^2r+L^2,$
\newline
$A^2=-\frac{1}{4}\Bigl(-2r\partial_r-3\Bigr)^2=(-4r\partial_rr\partial_r-9-12r\partial_r)\frac{1}{4}$
\newline
$=r\partial_r-r\partial_r^2r-\frac{9}{4}-3r\partial_r=-r\partial_r^2r-2r\partial_r-\frac{9}{4}$
;\newline so,\quad
$r^2(-\Delta)=-r\partial_r^2r+L^2=A^2+\frac{9}{4}+L^2-iA-\frac{3}{4}=A^2+L^2-iA-\frac{3}{4}.$\newline
(ii) \quad $\partial_\lambda\Bigl(e^{i\lambda A}|p|^\alpha
e^{-i\lambda A}\Bigr)=e^{i\lambda A}
i\Bigl[A,|p|^\alpha\Bigr]e^{-i\lambda A}=-\alpha e^{i\lambda A}
|p|^\alpha e^{-i\lambda A}$\newline $ \Rightarrow  e^{i\lambda A}
|p|^\alpha e^{-i\lambda A}=e^{-\lambda\alpha}|p|^\alpha .$ \newline
(iii)\quad As in(ii), but now, $i[A,r^\alpha]=\alpha r^\alpha .$
\newline (iv)\quad Follows from (v). \newline (v)\quad
$\Bigl\|\frac{1}{|p|}\frac{1}{r}\psi\Bigr\|_{L^2}
\leq2\Bigl\|r\frac{1}{r}\psi\Bigr\|_{L^2} = 2\|\psi\|$  by the
uncertainty inequality in 3 dimensions.$\blacksquare$
\enddemo
\proclaim {Proposition 2.2}

a) $\quad E_n(H) = E_n(H) E_n(|p|) E_n(H) + E_n(H) \delta^{-1}
O(2^{-n}) E_n(H).$

b) $\quad E_{\bar{n}}(|p|) E_n(H) = E_{\bar{n}}(|p|) O(2^{-n})
E_n(H), \quad \quad \bar{n} \neq n .$ \newline  For $\quad \eta > \ln
2,$

c) $\quad E_n (|p|) e^{i\eta A} E_n(H) = E_n(|p|) O(2^{-n}) E_n(H).$

d) $\quad \bar{E}_n (|p|) E_n(H) = \bar{E}_n(|p|) O(2^{-n}n) E_n(H).$
\endproclaim

\demo{Proof}
Part(a).

$E_n(H) = -E_n(E_n(|p|)-E_n)E_n + E_n(H)E_n(|p|) E_n(H),$

$E_n(|p|) - E_n(H) = \int \hat{E}_n(\lambda)\left(e^{i\lambda |p|} - e^{i\lambda H}\right) d\lambda$

$\quad \quad \quad \quad \quad \quad \quad \quad= \int \hat{E}_n(\lambda) e^{i\lambda H}
 \int_0^\lambda e^{-is H} (-i) V e^{is |p|} ds d\lambda .$
$$
    |V| \leq c<x>^{-2},
$$
$\Longrightarrow \left|\left( \phi, E_n\left[ E_n(|p|) - E_n(H)\right]E_n \psi \right) \right|$

$ \leq c \int \left\| <x>^{-1} e^{-is H} e^{i\lambda H} E_n \phi\right\| |\hat{E}_n(\lambda)\lambda|
 \frac{1}{\lambda} \int_0 ^\lambda \left\|<x>^{-1} e^{-is |p|} E_n \psi \right\| ds d\lambda$

$ \leq c\delta^{-1} \left\| HE_n\phi\right\| \int|\lambda \hat{F}_n(\lambda)|d\lambda \left\| |p|E_n \psi\right\|$

$ \leq c\delta^{-2} 2^{-n}2^n 2^{-n}\|\phi\| \|\psi\| =
c\delta^{-2}2^{-n}\|\phi\| \|\psi\|,$

where we used that,
$$\quad \quad <x>^{-1} \leq c|p| \leq c\delta^{-1}H \quad $$
in dimension three or higher, and
\newline
proposition(2.1).

Notice that,
\quad $\|pE_n\psi\| \leq c\delta^{-1}\|HE_n\psi\| \leq c\delta^{-1} 2^{-n}\|\psi\|$, \quad by Lemma 2.2.

Part(b).

$\left|\left(\phi,E_{\bar{n}} (|p|)E_n(H)\psi\right)\right| =
|\left(\phi,E_{\bar{n}}(|p|)\left[E_{\bar{n}}(|p|)-E_{\bar{n}}(H)\right]E_n(H)\psi\right)|$

$\leq  c\delta^{-1} \| HE_n\psi\| \||p|E_{\bar{n}}(|p|)\phi\|
\int|\hat{E}_{\bar{n}}\lambda|d\lambda ,$

by the proof of Part(a).
The last expression is therefore bounded by,
$$
    c \delta^{-1}2^{-n}\|\psi\|2^{-{\bar{n}}}\|\phi\|2^{\bar{n}} =
    c\delta^{-1}2^{-n}\|\phi\| \|\psi\|= O(2^{-n}).
$$

Part(c) follows from Part(b), since, for $|\lambda| > \ln 2$:
$$
\align
E_n(|p|)e^{i\lambda A} &= E_n(|p|)e^{i\lambda A} E_n(e^{-\lambda} |p|) \\
                        &\lesssim E_n(|p|) e^{i\lambda A} \sum_{|\bar n |\leq M} E_{\bar{n}} (|p|),\\
                        & \text {with} \quad \bar n \neq n, M<
                        \infty .
\endalign
$$

Part(d) follows from part(b), since the domain of $|p|$, in the
support of $\bar{E}_n(|p|)$, is covered by $n$ dyadic intervals from
$[2^{-n},1]. \blacksquare$
\enddemo

\proclaim{Proposition 2.3}
Assume that \quad $H\equiv|p|+V(x)$ \quad has no bound states or zero energy resonances,
and that $V(x)$ vanishes faster than $r^{-2}$ at infinity, and is sufficiently regular.
The dimension is 3. Then, for some $0<m<\infty$,
$$
H\geq m|p|. \tag 2.3
$$
\endproclaim
\demo{Proof}  Since $H$ has no bound states, and $V(x)
\rightarrow 0$ at infinity, $H\geq0$. If we now make a small
perturbation, $H\rightarrow H_\epsilon = H + \epsilon V,$  then,
since $H$ has no zero energy resonances, $H_\epsilon$ has no bound
states, for $\epsilon$ sufficiently small. Hence $H_\epsilon \geq
0$. \newline But then,
$$
    H=\frac{1}{1+\epsilon}\Bigl(|p|+(1+\epsilon) V \Bigr)+\Bigl(1-\frac{1}{1+\epsilon}\Bigr)
    |p|=\frac{1}{1+\epsilon}H_\epsilon +m|p|\geq m|p|.\blacksquare
$$
\enddemo

\proclaim{Lemma 2.3}
 For \quad $H=|p|+V$, we have that \quad
$|p|\geq \delta H$, for dimension$=3$, provided, \newline
$|V|\leq\bar{c}/r.$
\endproclaim
\demo{Proof}
 For $ \delta > 0$,
$$
\delta H=\delta|p|+\delta V \leq \delta|p| +\delta c |p| = \delta(1+c)|p| \leq |p| ,\tag 2.4
$$
where we used that for $|V|\leq\bar{c}|x|^{-1}$, we have $|V|\leq
c|p|$ in $3$ dimensions.\newline So, in particular, we have,
$$
    E_{I_n}(H)|p|E_{I_n}(H) \geq E_{I_n}(H)\delta HE_{I_n}(H)\geq \delta \inf I_n E_{I_n}(H). \tag 2.5
$$
$\blacksquare$
\enddemo
\proclaim{Proposition 2.4}
Suppose, as before, that $H\equiv |p|+V(x), $ the dimension is 3,
and that $V(x)$ is sufficiently regular, and vanishes faster than $r^{-2}$ at infinity.
 Suppose, moreover, that $H$ has no zero energy resonances, and no zero energy bound states. \newline
Then,
$$
    P_c(H)HP_c(H) \geq P_c(H)\delta |p| P_c(H),\quad \text{for some}\quad \delta>0 .\tag 2.6
$$
\endproclaim
\demo{Proof}
We have that,
$$
    P_c(H)\Bigl(|p|+V\Bigr)P_c(H)\geq 0.
$$
Add a small perturbation $\epsilon V$ to $H$: \quad $H_\epsilon
\equiv H+\epsilon V=|p|+(1+\epsilon)V .$ \newline Then, for $\epsilon$
sufficiently small, no new bound states are created. \newline Hence,
$$
    f(H_\epsilon\geq-\epsilon_0) \bigl(|p|+V+\epsilon V\bigr) f(H_\epsilon\geq -\epsilon_0)\geq 0 ,\tag 2.7
$$
for some $\epsilon_0 >0$, and $f$ is a smooth characteristic function of the interval $[-\epsilon_0 , \infty]$. \newline
Then,
$$
    P_c(H) H P_c(H) = P_c(H)\Bigl(\frac{1}{1+\epsilon}H_\epsilon+\epsilon_1 |p|\Bigr)P_c(H),
$$
would imply that:
$$
    P_c(H)HP_c(H) \geq \epsilon_1 P_c(H)|p|P_c(H),
$$
if we can prove that for some $\epsilon_2 < \epsilon_1$,
$$
    P_c(H)H_\epsilon P_c(H)\geq P_c(-\epsilon_2|p|)P_c .\tag 2.8
$$
So, want to use (2.7) to prove (2.8).
$$
\align
P_c (H)H_\epsilon P_c(H)&=P_c(H)f(H_\epsilon\leq-\epsilon_0)H_\epsilon P_c(H)
+P_c(H)f(H_\epsilon\geq -\epsilon_0)H_\epsilon P_c(H) \\
&\geq P_c(H)f(H_\epsilon \leq -\epsilon_0)H_\epsilon P_c(H), \tag 2.9
\endalign
$$
by (2.7).
Now,
$$
\align
P_c(H)f(H_\epsilon \leq-\epsilon_0)&=P_c(H)\Bigl[f(H_\epsilon \leq -\epsilon_0) - f(H\leq -\epsilon_0)\Bigr] \\
&= P_c(H)\int\hat{f}(\lambda)e^{+i\lambda H}i\int_0^\lambda e^{-isH}
\epsilon Ve^{isH_\epsilon}ds \\
&=P_c(H)O(\epsilon V)\int|\lambda  \hat{f}(\lambda)| d\lambda \\
&=P_c(H) O(\epsilon V) O(\epsilon_0^{-1}),
\endalign
$$
so,
$$
\align
&P_c(H)f(H_\epsilon\leq -\epsilon_0)H_\epsilon P_c(H) \\
& \quad = P_c(H)O(\epsilon V/\epsilon_0)\tilde{f}(H_\epsilon \leq-\epsilon_0)H_\epsilon
 \Bigl(f(H_\epsilon\leq-\epsilon_0)+\bar{f}(H_\epsilon \leq -\epsilon_0)\Bigr) P_c(H) \\
& \quad = P_c(H)O(\epsilon V/\epsilon_0)\tilde{f}(H_\epsilon)H_\epsilon f(H_\epsilon \leq -\epsilon_0)P_c(H) \\
& \quad = P_c(H)O(\epsilon
V/\epsilon_0)\tilde{f}(H_\epsilon)H_\epsilon O(\epsilon V/\epsilon_0)P_c ,\\
\endalign$$
since \quad $\tilde{f}(H_\epsilon\leq
-\epsilon_0)\bar{f}(H_\epsilon\leq -\epsilon_0) = 0.$
$(f+\bar{f}\equiv 1, \tilde{f} f=f)$.
 The last term, can be bounded by:
$$
    P_c(H)\epsilon^2<x>^{-2}O_{\epsilon_0}(H_{\epsilon})<x>^{-2}P_c(H)
     \leq O_{\epsilon_0}(1)P_c \epsilon^2|p|P_c ,\tag 2.11
$$
$$
O_{\epsilon_0} \sim O(1/\epsilon_0^2)H_{\epsilon}, \quad \text{coming
from :}
$$
$$
\int |\lambda \hat{f}(\lambda)|d\lambda = O(1/\epsilon_0).
$$
$\epsilon_0$ is basically the distance of zero to the largest
(negative) e.v. of $H_\epsilon$ , and $\epsilon$ is arbitrarily
small, so \quad $\epsilon^2 / \epsilon_0^2 \leq O(\epsilon^2)$.
Hence, (2.9) - (2.11) imply (2.8). \quad $\blacksquare$
\enddemo

\head  { Section 3. Maximal velocity bound} \endhead
 We begin with
estimating, for $a>1$,
$$
    E_n(H) F_a \left( \frac{|x|}{t} > a \right) E_n(H) \equiv E_n F_a E_n .\tag 3.1
$$
$$
    E_n F_a E_n = E_n F_a (F_n(A/t) + \bar {F}_n(A/t)) E_n, \tag 3.2
$$
where,
$$
    F_n(A/t) \equiv F_n(\frac{A}{t} > R2^{-n}) = 1 - \bar{F}_n ,\tag 3.3
$$
$$
    1 < R < a, \tag 3.4
$$
and such that
$$
 \bar{F}_n(2^{-n} b) F_a(b) = 0. \tag 3.5
$$

Since, by the propagation estimates,
$$
    \|F_n (A/t) E_n (H) \psi(t) \| \leq o(1) \|E_n\psi_0\|,
$$
with $o(1) \rightarrow 0$  as $t \rightarrow \infty$, we need to
control $E_n F_a \bar {F}_n E_n$ by a decaying function of $t$, in
order to prove the maximal velocity bound, on the energy shell
$I_n.$

\proclaim{Proposition 3.1}
$$
\|E_nF_aE_n\|\le o(1)\|E_n\psi\|+ \frac{c2^n}{t}\|E_n\psi\|
$$
\endproclaim
\demo{Proof}

$$
\align
E_n F_a \bar {F}_n E_n
&= E_n (|p|^{-1} |p|^{-1}\Delta) r^2 F_a r^{-2} \bar {F}_n E_n \\
&= E_n |p|^{-1} |p|^{-1} (A^2 + cA + c_1) (r^{-2} F_a) \bar {F}_n E_n \\
&= c_2 E_n |p|^{-1} (|p|^{-1} A r^{-1})(A+c) r^{-1} F_a \bar {F}_n E_n \\
& \quad + c_3 E_n |p|^{-1} (|p|^{-1} r^{-1}) r^{-1} F_a \bar {F}_n
E_n \equiv B_1 + B_2. \tag 3.6
\endalign
$$

$$
B_2 = E_n O(2^n)O_0(1)c_3
\frac{1}{at}\left[\left(\frac{ta}{r}\right)F_a\right]\bar{F}_n E_n,
\tag 3.7
$$
where we used that,
$$
E_n |p|^{-1} = O (2^{+n}) \quad \text{and} \quad |p|^{-1}r^{-1} =
O_0(1).
$$
Since \quad $\frac{ta}{r} F_a = O(1) , \quad B_2 = O(2^n/at),$
$$
\align
B_1 &= E_n O(2^n) O_1(1) c \frac{1}{at} \left( \frac{ta}{r} F_a\right) \bar {F}_n E_n \\
&\quad + E_n O(2^n) O_1(1) \frac{1}{at} \left(\frac{ta}{r}F_a \right) Rt (Rt)^{-1} A \bar {F}_n E_n \\
&\quad + E_n O(2^n) O_1(1) \frac{1}{at} \left[A, \left(\frac {at}{r}
F_a \right) \right] \bar {F}_n E_n,
\endalign
$$
where \quad $$O_1(1) \equiv |p|^{-1}( A-3/2\imath)
r^{-1}=|p|^{-1}p\cdot xr^{-1}=\sum_{k=1}^3
\frac{p_k}{|p|}\frac{r_k}{r}.$$ Therefore,
$$
\align
B_1 &= E_n O(2^n / at) \bar{F}_n E_n + E_nO(2^n)O_1(1)\left(\frac{ta}{r}F_a\right) \frac{R}{a}O(2^{-n})\bar{F}_n E_n\\
&= O(2^n/at) + O\left(\frac{R}{a}\right). \tag 3.8
\endalign
$$
The key to this computation is the repeated use, as we do below, of the following:
$$
\align
E_n &= E_n|p|^{-1}|p|^{-1}(A^2+cA+c_1) r^{-2} \\
&= c_2 E_n|p|^{-1}(|p|^{-1}Ar^{-1})(A+c^{'})r^{-1} + c_3
E_n|p|^{-1}(|p|^{-1}r^{-1})r^{-1}\\
&=E_n|p|^{-1}O_1(1)(A+c^{\prime})r^{-1}+c_3
E_n|p|^{-1}(|p|^{-1}r^{-1})r^{-1}\\
&+c_4E_n|p|^{-1}(|p|^{-1}r^{-1}A)r^{-1}. \tag 3.9
\endalign
$$
We apply it again, to the \quad $ O\left(\frac{R}{a}\right)$ \quad
term:$ (F^{[k]}(\lambda)\equiv \lambda^k F(\lambda ));$
$$
\align
O(R/a) \sim &\frac{R}{a}E_n O_1(1)F_a^{[-1]}\bar{F}_n^{[1]} E_n\\
&= \frac{R}{a}E_n|p|^{-1}O_1(1)(A+c^{'})r^{-1}O_1(1)F_a^{[-1]}\bar{F}_n^{[1]}E_n \\
& \quad + \frac{R}{a}E_n|p|^{-1}O_0(1)r^{-1}O_1(1)F_a^{[-1]}
\bar{F}_n^{[1]}E_n. \tag 3.10
\endalign
$$
Now, the important observation is that,
$$
 r^{-1}|p|^{-1}Ar^{-1} = (r^{-1}|p|^{-1}A)r^{-1}
$$
and
$$
r^{-1}O_1(1)
=O_1(1)r^{-1}-r^{-1}O(|p|^{-1})r^{-1}=O_1(1)r^{-1}+O(1)r^{-1}
$$
$$
\quad
[A,O_1(1)] \sim O_1(1). \tag 3.11
$$
We derive:
$$
O_1(1)(A+c^{'}) r^{-1} O_1(1) F_a^{[-1]} \bar{F}_n^{[1]}=O_1(1)(A+
c^{'})(r^{-1}p^{-1}Ar^{-1})F_a^{[-1]}\bar{F}_n^{[1]},
$$
$$
\align Ar^{-1}F_a^{[-1]}\bar{F}_n^{[1]}
&=r^{-1}F_a^{[-1]}A\bar{F}_n^{[1]}+cr^{-1}F_a^{[-1]}\bar{F}_n^{[1]}\\
&+\frac{1}{at}(\partial_r F_a^{[-1]})\bar{F}_n^{[1]}\\
&=\frac{1}{at} F_a^{[-2]}(Rt)2^{-n}\bar{F}^{[2]}
+\frac{c}{at}F_a^{[-2]}\bar{F}_n^{[1]}+\frac{1}{at}(\partial_{\lambda}F_a^{[-1]})
\bar{F}_a^{[1]}\\
&=O(\frac{1}{at})+O(2^{-n}\frac{R}{a}F_a^{[-2]}\bar{F}_n^{[2]}).
\endalign
$$
$$
\partial_\lambda F(\lambda=x)\equiv\frac{\partial F}{\partial
\lambda}\bigg|_{\lambda=x}.
$$
Using that $ A r^{-1} p^{-1}=O_1(1)+O_0(1)$, we have,
$$
\align \frac{R}{a}E_n O_1(1)F_a^{[-1]}\bar{F}_n^{[1]}E_n
&= \frac{R}{a}E_n|p|^{-1} O_1(1)(r^{-1}|p|^{-1}A+c^{'})r^{-1}F_a^{[-1]}\bar{F}_n^{[1]}E_n  \\
& \quad + \frac{R}{a}E_n|p|^{-1}O_{1}(1)^2 r^{-1} F_a^{[-1]}A\bar{F}_n^{[1]}E_n \\
&= O\left(\frac{R}{a}\frac{2^n}{at}\right) + E_n\left(\frac{R}{a}
\right)^2 O_{0,1}(1)^2 F_a^{[-2]} \bar{F}_n ^{[2]} E_n .\tag 3.12
\endalign
$$

 where we used that,
$$
    E_n |p|^{-1} = E_n O (2^{-n}), \quad \quad r^2 \Delta = A^2 + cA + c^{'},
$$
$$
    \frac{A}{R2^{-n}t} \bar {F}_n = O(1),
$$
$$
    |p|^{-1} A r^{-1} = O(1).
$$
We now do this computation again, this time, for the term of order
$(\frac{R}{a})^2$: \newline Doing it $k$ times, we get:
$$
    \left(\frac{R}{a}\right)^k E_n O(1)^k F_a^{[-k]} \bar{F}_n^{[k]} E_n + \sum_{j=1} ^k E_n O\left(\frac{2^n}{t}
    \right) \left(\frac{R}{a}\right)^{(j-1)} \frac{1}{a} O(1)^j \tilde{F}_a^{[-j]} \bar{F}_n^{[j-1]}E_n ,\tag 3.13
$$
with,
$$
\tilde{F}_a^{[-k]} = \frac{r}{at} F_a^{[-k+1]} + r\partial _r
\frac{r}{at} F_a ^{[-k+1]}. \tag 3.14
$$

For \quad $k \sim \delta \ln t,$ \quad and \quad $\frac{R}{a} < 1$ \quad sufficiently small,
 we have that the  $E_n F_a \bar{F}_n E_n$ \quad term is bounded by \quad
 $O(t^{-1}2^n)$.$\blacksquare$
 \enddemo
 We can now prove the Theorem on Maximal velocity bound.
\smallskip
 {\bf Proof of Maximal Velocity Bound}
\medskip

Now,
$$
\align
&\left(\psi (t), F_a^2 (\frac{r}{t}>a) \psi(t) \right) \\
&\quad \quad=\sum_n \left(H^{-1/2} \psi(t), F_a ^2 E_n H^{1/2} \psi(t)\right) + Q \\
&\quad \quad= \sum_n \left( H^{-1/2} \psi(t), F_a ^2(F_n  (A/t) + \bar{F}_n (A/t)) E_n H^{1/2} \psi(t) \right) + Q \\
&\quad \quad= \sum_n ( H^{-1/2}\psi, F_a^2F_nE_n H^{1/2}\psi) +
\sum_n(H^{-1/2}\psi, F_a^2\bar{F}_nE_nH^{1/2}\psi) + Q .\tag 3.15
\endalign
$$

$$
\align
&|\sum_n(H^{-1/2}\psi,F_a^2F_nE_n H^{1/2} \psi)| \\
&\quad \quad \quad \quad \leq \|F_a H^{-1/2} \psi\| \sum_n<n>^{-1/2-\varepsilon/2} <n>^{1/2+\varepsilon/2}
 \|F_n E_n H^{1/2} \psi\| \\
&\quad \quad \quad \quad \leq c\|F_a H^{-1/2} \psi \|
 \left( \sum_n <n>^{1+\varepsilon} \| F_n E_n H^{1/2}
 \psi\| ^2 \right) ^{1/2}\\
&\quad \quad \quad \quad \leq c\|F_a H^{-1/2} \psi\|
 \left(\sum_n o(1) \| H^{1/2} E_n <x>^{1/2} \psi \|^2 <n>^{1+\varepsilon }\right)^{1/2} \\
&\quad \quad \quad \quad \leq c\|F_a H^{-1/2} \psi\| o(1)
\|<x>^{1/2} \psi \|, \tag 3.16
\endalign
$$

$$
\align
\left|\sum_n ( H^{-1/2} \psi, F_a^2 \bar{F}_n E_n H^{1/2} \psi)\right|
&\leq \left|\sum_n (F_a H^{-1/2} \psi, \bar{F}_n F_a E_n H^{1/2} \psi)\right|\\
& + \left|\sum_n(F_a H^{-1/2} \psi, O(\frac{1}{at})(R2^{-n})^{-1} O(1) E_n H^{1/2} \psi) \right| \\
& \leq c\left\|F_a H^{-1/2} \psi \right\| \sum_n \left\|2^{+n/2} E_n
\psi \right\| \frac{1}{t}\frac{R}{a}. \tag 3.17
\endalign
$$
It follows that,
$$
      \left(\psi(t),F_a^2 \psi(t)\right) \leq c\left\|F_a H^{-1/2} \psi_0 \right\|
      \quad \left\| |\ln H|^{\frac{1+\varepsilon}{2}} H^{-1/2} <x>^{1/2} \psi_0 \right\| o(1) + Q ,\tag 3.18
$$
$$
    Q \equiv \left( H^{-1/2} \psi(t), \left[H^{1/2}, F_a^2\right] \psi(t)\right). \tag 3.19
$$
To control Q, (3.19), we need to commute fractional powers of $H$.
To this end we use that:
$$ H^\alpha = c_\alpha \int_0^\infty \frac{\lambda^{\alpha-1}}{\lambda+H}Hd\lambda, $$
and estimate,
$$
\align
\lambda^{\alpha -1}\left[\frac{H}{H+\lambda},F_a\right]
&= \lambda^{\alpha -1} H \left[\frac{1}{H+\lambda},F_a \right] + [H, F_a]\frac{\lambda^{\alpha -1}}{H+\lambda} \\
&= \frac{-H}{H+\lambda} O\left(\frac{1}{t}F_a^{\prime}\right)
\frac{\lambda^{\alpha -1}}{H+\lambda} + O\left(\frac{1}{t}F_a^{\prime}\right)
\frac{\lambda^{\alpha -1}}{H+\lambda} \equiv \circledast\\
 \int_0 ^\infty d\lambda \circledast &= O(1) O\left(\frac{1}{t}F_a^{\prime}\right)O(H^{-\alpha}). \tag 3.20
\endalign
$$
Therefore, using (3.20) with $\alpha=1/2,$we have that

$$
|Q|=|(H^{-1/2}\psi(t),O(1)O(\frac{1}{t})H^{-1/2}\psi(t))|\le\frac{c}{t}\|H^{-1/2}\psi(t)\|^2
\le\frac{c}{t}\|<x>^{1/2}\psi(0)\|^2.
$$
End of Proof.

\newpage

{\bf REFERENCES}
\medskip
[BFS]~  V. Bach, J. Fr\"ohlich, and IM Sigal.: Commun. Math. Phys.,
{\bf 207} :249-290, 1999

\medskip

[BFSS]~  V. Bach, J. Fr\"ohlich, I. M. Sigal, A. Soffer, {\it
Positive Commutators and Spectrum of Pauli-Fierz Hamiltonians of
Atoms and Molecules}, Comm. Math. Phys., {\bf 207}, 1999, 557-587.
\medskip
[FGS]~ J. Fr\"ohlich, M. Griesemer, I. M. Sigal {\it Spectral
renormalization and Local Decay...}, (2009), arxiv0904.1014v1

\medskip
 [HSS]~ W. Hunziker, I. M. Sigal, A. Soffer, {\it Minimal
Velocity Bounds}, Comm. PDE, {\bf 24}, (1999), No. 11/12, 2279-2295.
\medskip

[Sig-Sof]~I.M. Sigal and A. Soffer, {\it Local Decay and Propagation
Estimates for Time Dependent and Time Independent Hamiltonians}, Preprint, Princeton 1988,
(ftp://www.math.rutgers.edu/pub/soffer).

\enddocument

\end